\documentclass{article}

\usepackage{arxiv}

\usepackage[utf8]{inputenc} 
\usepackage[T1]{fontenc}    
\usepackage{hyperref}       
\usepackage{url}            
\usepackage{booktabs}       
\usepackage{amsfonts}       
\usepackage{nicefrac}       
\usepackage{microtype}      
\usepackage{lipsum}

\usepackage{graphicx}
\usepackage{amsmath,amssymb}
\pdfoutput=1
\usepackage{tabularx}
\usepackage{changepage}

\usepackage{textcomp,marvosym}
\usepackage[right]{lineno}
\usepackage{microtype}
\DisableLigatures[f]{encoding = *, family = * }
\usepackage[table]{xcolor}
\usepackage{array}

\DeclareUnicodeCharacter{2212}{-}

\newcolumntype{+}{!{\vrule width 2pt}}

\newlength\savedwidth

\makeatletter

\title{Double stochastic opinion dynamics with fractional inflow of new opinions}

\author{
  Vygintas~Gontis\thanks{Use footnote for providing further
    information about author (webpage, alternative
    address)---\emph{not} for acknowledging funding agencies.} \\
  Institute of Theoretical Physics and Astronomy\\
  Vilnius University\\
  Saul{\. e}tekio al. 3, 10257 Vilnius, Lithuania \\
  \texttt{vygintas@gontis.eu} \\
}

\begin{document}
\maketitle

\begin{abstract}
A recent analysis of empirical limit order flow data highlights the necessity for a more refined order flow model that integrates the power-law distribution of limit order cancellation times. These cancellation times follow a discrete probability mass function derived from the Tsallis $q$-exponential distribution, or equivalently, the second form of the Pareto distribution. By combining fractional L'{e}vy stable motion as the model for limit order inflow with the power-law distribution for cancellation times, we propose an innovative approach to modeling order imbalance in financial markets. We extend this model to a broader context, illustrating its applicability to opinion dynamics in social systems where opinions have a finite lifespan. This proposed model exemplifies a stochastic time series characterized by stationary increments and broken self-similarity. Consequently, it offers a novel framework for testing methods to evaluate long-range dependence in such time series.
\end{abstract}

\keywords{Time-series and signal analysis \and Discrete, stochastic dynamics \and Scaling in socio-economic systems \and Fractional dynamics \and Quantitative finance}

\section{Introduction}

The debate in the scientific community of power-law behavior in social and physical systems is long-lasting \cite{Newman2005CPh,Kumamoto2018FPh}. Usually, one observes power-law behavior at the macro level of the system and looks for the microscopic interpretation of observed phenomena. From the mathematical point of view, the power-law is the only function satisfying the scale-free property $p(b x)=f(b)p(x)$ \cite{Newman2005CPh}. Thus, there is a close relation between the self-similarity of stochastic processes and power-law \cite{Newberry2019PhysRevLett}. Power-law statistical property is a characteristic feature of social and financial systems. The measures of long-range memory based on self-similarity are ambiguous as Markov processes with power-law statistical properties can exhibit long-range memory, including slowly decaying auto-correlation \cite{Gontis2004PhysA,McCauley2006PhysA,Gontis2006JStatMech,McCauley2007PhysA,Gontis2008PhysA,Micciche2009PRE,Micciche2013FNL,Ruseckas2011PRE,Kononovicius2015PhysA}. The financial markets provide us with empirical limit order book (LOB) data that exhibit power-law statistical properties as well \cite{Gould2013QF}.

From the econophysics perspective, it is preferable to provide the microscopic interpretation of the econometric models, primarily serving as macroscopic descriptions of intricate social systems. These models often rely on self-similarity and long-range dependence assumptions. To advance our comprehension of long-range memory in social systems, it becomes imperative to juxtapose macroscopic modeling with empirical analyses. In our prior review~\cite{Kazakevicius2021Entropy}, we raised the question of whether the observed long-range memory in social systems results from genuine long-range memory processes or is merely an outcome of the non-linearity inherent in Markov processes.

In this contribution, we demonstrate how vital the assumptions of fractional L'{e}vy stable motion (FLSM) are and how straightforward the model of opinion dynamics breaks these assumptions. The proposed model is empirically grounded on the order disbalance time series of the financial markets \cite{Gontis2023FractFrac,Gontis2022CNSNS}. As recorded in order books, market-order flows exhibit long-range persistence, attributed to the order-splitting behavior of individual traders~\cite{Lillo2005PhysRevE}. This discovery reinforces the presence of genuine long-range memory in financial systems, as recently confirmed in a comprehensive investigation~\cite{Sato2023PhysRevLett}. The order-splitting behavior of individual traders should be discernible in the sequence of submitted limit orders.

Section \ref{OpinDynamic} presents a short description of the limit order time series serving as the background for the more general interpretation of opinion dynamics. In Section \ref{WaitingTime}, we present a model of power-law waiting time originating from the system of heterogeneous agents, and in Section \ref{Self-Similarity}, we provide evidence of broken self-similarity assumption when cancellation of opinions is included in the model. We discuss our results and provide conclusions in Section \ref{sec:Conclusions}.

\section{Modeling limit order flow and/or opinion dynamics \label{OpinDynamic}}

In our work~\cite{Gontis2023FractFrac}, we delved into the sequence of limit order submissions to the market, denoted as $X_L(j)$,
\begin{equation}
X_L(j)=\sum_{i=1}^{j} v_i = \sum_{i=1}^{j} Y_L(i),
\label{eq:limit-order-sequence}
\end{equation}
where $v_i$ represents the volume of the submitted limit order. We examined the series $X_L(j)$ through the lens of FLSM, as the probability density functions (PDF) of order volumes $v_i$ have power-law tails, and we documented fluctuations of the memory parameter for various stocks in the region $d \simeq 0.19\div0.34$. Despite the rough approximation of the PDF of volumes $v_i$ by the L'{e}vy stable distribution, the time series $X_L(j)$ can be considered FLSM-like.

The series $X_L(j)$ functions as a macroscopic measure of opinion in the order flow and exhibits long-range dependence due to the heterogeneity of agents. Nevertheless, it is more prudent to consider the measure of traders' macro opinion, incorporating events of order cancellation and execution. Therefore, we explore an alternative sequence of order flow
\begin{equation}
X(j)=\sum_{i1\leqslant j < i2}v_{i1,i2}=\sum_{i=1}^{j} Y(i),
\label{eq:order-disbalance}
\end{equation}
where the first sum is over all live limit orders, including all limit order volumes $v_{i1,i2}$ submitted before event $j$ and waiting for cancellation or execution. A sequence of limit order submissions of length $N$ generates a series of order disbalance $X(j)$ of length $2N$ since each submission is paired with a cancellation or execution event. Notably, series $X(j)$ displays a few crucial differences from series $X_L(j)$. Firstly, the empirical sequence $X(j)$ appears bounded, while $X_L(j)$ is unbounded. Secondly, we obtain contradictory results when evaluating the memory parameter $d$ using the assumption of FLSM for the series $X(j)$. In our previous work~\cite{Gontis2023FractFrac}, we concluded that the time series defined in \eqref{eq:order-disbalance} as order disbalance is not FLSM-like. Consequently, the persistent limit order submission flow or long-range dependence is concealed from econometric methods defining memory in the time series of order disbalance $X(j)$.

In the quest for a new interpretation of order disbalance series $X(j)$, we have introduced the concept of a discrete $q$-exponential probability mass function
\begin{equation}
P_{\lambda,q}(k)=SP_{\lambda,q}(k-1)-SP_{\lambda,q}(k)=(1 + (q-1) (k-1) \lambda)^{\frac{2 - q}{1 - q}} - (1 + (q-1) k \lambda)^{\frac{2 - q}{1 - q}},
\label{Eq:PMFfromSF}
\end{equation}
as a $q$-extension of the geometric distribution, grounded in the theoretical foundations of generalized Tsallis statistics~\cite{Tsallis1988-ku}. This distribution allows for a more accurate fit of empirical limit order cancellation times, revealing their weak sensitivity to order sizes and price levels. The fitted discrete $q$-exponential PMF parameters, $\lambda=0.3$, and $q=1.5$, remain consistent across ten stocks and trading days analyzed. The power-law distribution of cancellation (waiting) times might originate from a stochastic queueing model in which tasks are executed according to a continuous-valued priority \cite{Barabasi2005Nature,Grinstein2008PRE}. Instead, in Section \ref{WaitingTime}, we derive the power-law of waiting time originating from the heterogeneity of agents.

We propose a relatively straightforward limit order flow disbalance model by combining fractional L'{e}vy stable limit order inflow with the $q$-exponential lifetime distribution. This model arises as an interpretation of limit order flow empirical analysis considered in \cite{Gontis2023FractFrac} and is an illustrative example of broader modeling of social systems.

Let us generalize the interpretation of the proposed model, considering it as a possible version of the opinion dynamics, probably applicable to other social systems as well. In its original interpretation, the model involves two random sequences:
(a) A sequence of limit order volumes $v_i$ generated as ARFIMA\{0,d,0\}\{$\alpha,N$\}, where $d$ is the memory parameter, $\alpha$ is the stability index, $N$ is the length of the sequence.
(b) A corresponding independent sequence of limit order cancellation times with the same length $N$ generated using the probability mass function (PMF) $P_{\lambda,q}(k)$ defined by Equation~\eqref{Eq:PMFfromSF}.

For the extended interpretation of the model, we assume that every $v_i$ is the opinion weight, positive for the first of two possible (buy) and negative for the second (sell). The limit order cancellation time measured in the event space $k=i2-i1$ will mean opinion lifetime. Though the model is proposed for analyzing limit order flow in the financial markets, its extended interpretation might help investigate other cases of weighted opinions in social systems. In this study, the proposed model is helpful as an example of a relatively simple time series constructed using the ARFIMA sequence but exhibiting properties outside the assumption of self-similarity. 

With these two independent sequences, we can calculate the model time series $X(j)=\sum_{i=1}^{i=j}Y(i)$ defined by sequence $v_{i1,i2}$, see Equation~\eqref{eq:order-disbalance}. Here, opinion submission event number $i1$ and its cancellation event number $i2$ can be calculated for every $v_i$ of sequence (a) and corresponding $k$ of sequence (b). The generated random sequence represents the artificial analog of order disbalance time series used to compare with empirical order flow in the financial markets \cite{Gontis2023FractFrac}. We achieved good correspondence of the artificial model with empirical data, choosing parameters of the artificial model as follows: $\alpha=1.8$; $\lambda=0.3$; $q=1.5$, \cite{Gontis2023FractFrac}. Note that for other applications, the model can be simplified replacing the sequence of $v_i$ with unit weights
\begin{equation}
vu_i=Sign(v_i)=\begin{cases}-1 & \text { if } v_i<0 \\ 0 & \text { if } v_i=0 \\ 1 & \text { if } v_i>0\end{cases}.
\label{eq:Sign-volume}
\end{equation}
We will denote these series with an additional index of $S$, for example, $X_S(j)$. One more direction of possible simplification would be the choice of $q=1$ in Equation~\eqref{Eq:PMFfromSF}, giving us a Geometric distribution
\begin{equation}
\lim_{q \to 1} P_{\lambda,q}^{(ds)}(k) = \exp^{-y \lambda} (\exp^\lambda-1) = (1-p)^{k-1} p,
\label{Eq:q-limit}
\end{equation}
where we denote $p=1-\exp^{-\lambda}$. The Geometric distribution as a discrete version of the exponential distribution is the most common choice for the waiting time in many physical and social systems.

\section{Heterogeneity of agents and power-law of waiting time \label{WaitingTime}}

\begin{figure}[h]
\begin{centering}
\includegraphics[width=0.5\textwidth]{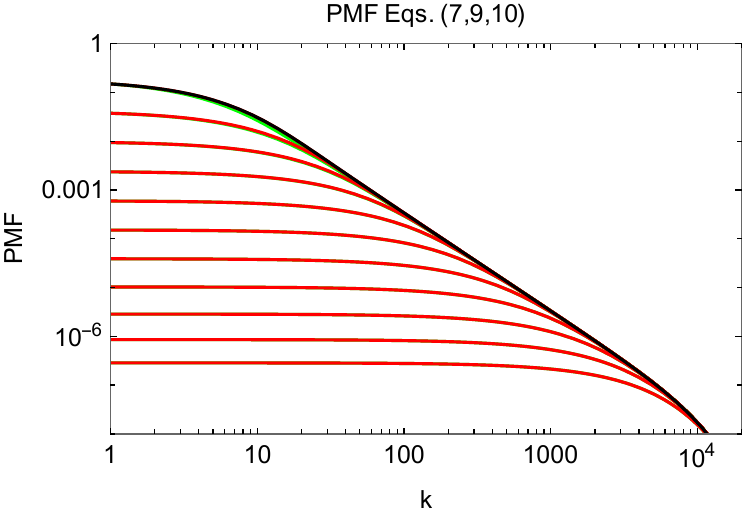}
\end{centering}
\caption{Visualization of PMF's Eqs. \eqref{Eq:PMF-ensemble}, \eqref{Eq:PMF-Geom}, \eqref{Eq:PMF-Exp}. Black line: represents PMF \eqref{Eq:PMF-ensemble}; green line: \eqref{Eq:PMF-Geom}; and red line: \eqref{Eq:PMF-Exp}.  \label{fig1}}
\end{figure}

Though, power-law waiting time appears in the stochastic queueing model with a continuous-valued priority \cite{Barabasi2005Nature,Grinstein2008PRE}, one more reasonable explanation of the power-law could be the heterogeneity of agents. Trading agents in the financial markets have very different assets in their disposition. Thus, the trading activity of agents varies on a wide scale, and the lifetime of their orders is very different. We can propose a simple approach to combining the heterogeneity of agents seeking to derive the power-law distribution of limit order cancellation time. Assume we have $n$ categories of agents with different limit order submission (cancellation) rates. The lowest rate is one limit order per trading day (duration of time series investigated). Let us denote this probability as $0<p_1=\eta/n<1$. Then agents, who submit two orders per day, have twice higher probability, and agents who submit $i$ limit orders are characterized with probability $p_i=i \eta/n$. The most active trader submits $n$ limit orders with probability $p_n=\eta$. Continuing with such definition, we have an individual Geometric distribution of the waiting (cancellation) time for the $i$-th category of agents $P_i(k)=(1-p_i)^{k-1} p_i$. Finally, we have to average this PMF over all categories of agents. It is clear from our assumption that the probability of order arrival from different categories of agents might be equal as agent order submission frequency is proportional to the index $i$ and the number of agents in a category is inversely proportional -- Zipf's law. Thus, we can write the PMF of order (opinion) waiting time for the whole ensemble of agents as follows
\begin{align}
P_{\eta,n}(k) & = \sum_{i=1}^n(1-\frac{\eta i}{n})^{k-1} \eta i/n^2\simeq \frac{\eta}{n^2}\sum_{i=1}^n i\exp(-\frac{\eta(k-1)i}{n}) \label{Eq:PMF-sumGE} \\& =\frac{\eta (n \exp(-\eta (k-1)) - (1+n)\exp(-\eta (k-1)+\frac{\eta (k-1)}{n})+\exp(\frac{\eta (k-1)}{n}))}{n^2 (1-\exp(\frac{\eta (k-1)}{n}))^2}.
\label{Eq:PMF-ensemble}
\end{align}
Seeking to clarify the result Eq. \eqref{Eq:PMF-ensemble} one can find the limit 
\begin{equation}
\lim_{n\rightarrow \infty} P_{\eta,n}(k)=\frac{1-\exp(-\eta (k-1)) (1+\eta (k-1))}{\eta (k-1)^2},
\label{Eq:PMF-limit}
\end{equation}
giving the evidence of the power-law with the exponent $\kappa=2$. We can recover the power-law nature of PMF Eq. \eqref{Eq:PMF-ensemble}  plotting it in Fig. \ref{fig1} together with partial sums
\begin{align}
P_{\eta,n,m}^{Geom}(k) & = \frac{\eta}{n^2}\sum_{i=1}^{2^m} i (1-\frac{\eta i}{n})^{k-1} \label{Eq:PMF-Geom}\\ P_{\eta,n,m}^{Exp}(k) & = \frac{\eta}{n^2}\sum_{i=1}^{2^m} i\exp(-\frac{\eta(k-1)i}{n}),
\label{Eq:PMF-Exp}
\end{align}
where we use the set of $m=\{0,1,2....10\}$ and $n=2^{10}=1024$.
Note that green and red lines are indistinguishable as a result of Eqs. \eqref{Eq:PMF-Geom} and \eqref{Eq:PMF-Exp} coincide. The black line coincides with both partial sums when $m=10$.
One can note that our assumptions incorporating Zipf's law lead to the power-law of cancellation (waiting) time exponentially stretched on both sides. This natural restriction comes from the fixed number of agent categories $n$ or a related number of opinions (orders) submitted $N=n (n+1)/4$. The most important finding is that the power-law exponent in Eq. \eqref{Eq:PMF-ensemble} is $\kappa=2$. From the relation of $q$-exponential distribution with Pareto distribution, we can easily find that related exponent $q=1.5$ as we have defined empirically in \cite{Gontis2023FractFrac}. Thus, the presented description of PMF of waiting time for the ensemble of heterogeneous agents strengthens the conclusion that power-law exponent $q=1.5$ is a stylized fact of the financial markets. A more detailed investigation of empirical cancellation times using proposed PMF \eqref{Eq:PMF-ensemble} would be helpful. 

\section{Self-similarity analysis of proposed model \label{Self-Similarity}}

Continuing our efforts to understand long-range memory in social systems \cite{Kazakevicius2021Entropy}, it becomes essential to compare the macroscopic description with empirical analyses and agent-based modeling. The empirical investigation of volatility, trading activity, and order flow in the financial markets has provided solid ground for empirical investigations of long-range memory properties \cite{Baillie1996JE,Engle2001QF,Plerou2001QF,Gabaix2003Nature,Ding2003Springer,Lillo2005PhysRevE,Sato2023PhysRevLett}. Various econometrical models with fractional noise have been proposed to describe volatility time series \cite{Ding1993JEmpFin,Baillie1996JE,Bollerslev1996Econometrics,Giraitis2009,Conrad2010,Arouri2012,Tayefi2012}. However, from the perspective of econophysics, these models primarily serve as macroscopic descriptions of complex social systems, often based on ad hoc assumptions of long-range memory. As a result, despite applying advanced trading algorithms and machine learning techniques, predicting stock price movements remains challenging for researchers \cite{Alec2015QF,Kumar2018IEEE,Zaznov2022Mathematics}.

Here, we will demonstrate that the requirement of self-similarity, widely used in modeling long-range dependence, is firm and challenging in the proposed opinion dynamics model. Usually, the econometric methods are used without questioning the assumption of self-similarity for the time series investigated; see \cite{Gomez2022FinInnov} for a more detailed consideration of the problem. 

The theory of stochastic time series is based on self-similar processes with stationary increments; thus, we need to recall these concepts. Stochastic time series $X(t)$ can be considered as self-similar if the scaling relation between two distributions is fulfilled $X(\tau t) \sim \tau^H X(t)$, here $\sim$ means that two distributions are the same for any $\tau>0$ and $t>0$. One more requirement is stationarity of increments $X(t + \tau) − X(t) \sim X(\tau ) − X(0)$ for any $\tau>0$ and $t>0$. A self-similar process with stationary increments has self-affine increments $X(t + c \tau ) − X(t) \sim \tau^H (X(t + c) − X(t))$ for any $c>0$, \cite{TrinidadSegovia2012PhisA}. All these properties are defined using equality in distributions; thus, the most straightforward estimation of $H$ should also be based on the equality in distributions. By testing the distributions, we can discriminate cases when series having stationary increments deviate from the requirement of self-similarity. 

It is obvious that in the proposed modeling of limit, order flow $X_L(j)$ and opinion disbalance $X(j)$ increments are stationary as they originate from the L'{e}vy stable distribution. Following the definitions above, we can write the condition of self-similarity  as
\begin{equation}
|X(t + \tau) − X(t)| \sim \tau^H |X(1) − X(0)|.
\label{Eq:self-similarity}
\end{equation}
For the comparison of distributions, we use Kolmogorov Smirnov (KS) two-sample test \cite{Hodges1958Matematik} and calculate KS distance $D$
\begin{equation}
D=sup_{x}|F_{\tau_1}(x) − F_{\tau_k}(x)|,
\label{Eq:KS-distance}
\end{equation}
where cumulative empirical distribution functions $F_{\tau_i}(x)$ for the integer sequence of $i={0,1,2,...}$ and corresponding sequence of $\tau_{i,H}=2^i$ is defined as
\begin{equation}
F_{\tau_i,H}(x) = P[|\frac{X(t + \tau) − X(t)}{\tau^H}| \leq x].
\label{Eq:Empirical-CDF}
\end{equation}
From the definition of self-similarity \eqref{Eq:self-similarity}, it follows that we should get the same $H$ minimizing distance \eqref{Eq:KS-distance} for any $\tau$. If one gets various values of $H$ for different $\tau$ values, the requirement of self-similarity is not fulfilled.

The proposed model of opinion dynamics is a good example of a time series to demonstrate how self-similarity is broken when opinion cancellation is introduced into a self-similar series, FLSM, of opinion inflow $X_L(t)$. We generate FLSM series $X_L(j)$ with parameters: $\alpha=1.8$, $d=0.3$, $N=200 000$ and corresponding series of opinion duration (waiting time) $k(j)$  using \eqref{Eq:PMFfromSF} with parameters $q=1.5$ and $\lambda=0.3$. Then we calculate series $X(j)$ of length $2N$ of opinion disbalance \eqref{eq:order-disbalance}.
\begin{figure}[h]
\begin{centering}
\includegraphics[width=0.9\textwidth]{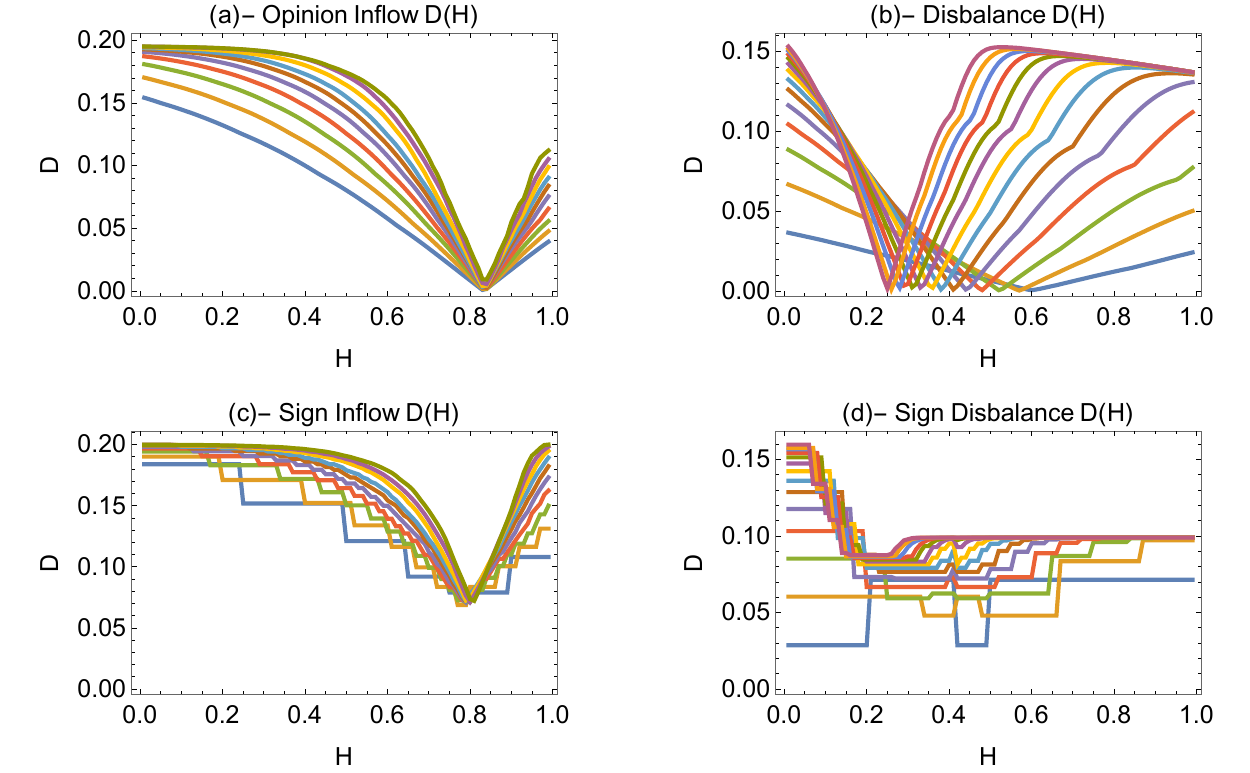}
\par\end{centering}
\caption{Numerical KS distance as a function of $H$. (a) for FLSM inflow of opinions $X_L(j)$; (b) for opinion disbalance $X(j)$ including opinion cancellation events; (c) for the simplified inflow of signs $X_{S,L}(j)$; (d) for the disbalance with the simplified inflow of signs $X_S(J)$. Model series generated using parameters $d=0.3$, $\alpha=1.8$, $\lambda=0.3$, $N=200000$. \label{fig2}}
\end{figure}
In Fig. \ref{fig2} we compare numerically calculated KS distances $D(H)$, Eq. \eqref{Eq:KS-distance}, as functions of $H$ for the series $X_L(t)$, sub-figure (a); for the series $X_(j)$, sub-figure (b); for the series $X_{S,L}(j)$, subfigure (c); for the series $X_S(j)$ . Time series $X_L(j)$ is self-similar as follows from its definition, and series $X_(j)$ is not self-similar as we get the set of various $H$ values for the different values of $\tau_i=2*2^i$. Though in the case of the simplified model taking only signs of volumes, see Eq, \eqref{eq:Sign-volume}, KS distance $D(H)$ is less sensitive to $H$, numerical results confirm that we can consider series $X_{S,L}(j)$ as self-similar, and series $X_S(j)$ are not self-similar. From our point of view, this procedure to control self-similarity should apply to any observed time series.

Nevertheless, we must admit that researchers use various methods to estimate the self-similarity parameter $H$ of observed time series without testing the self-similarity requirement itself \cite{Gomez2022FinInnov}. From our point of view, we must pay much more attention to developing new methods to test self-similarity assumptions. The method we proposed here appears less accurate for the simplified series $X_{S,L}(j)$ and $X_S(j)$, as numerically calculated functions $D(H)$ reveal some fractured structure. 
\unskip 
\begin{table}[h] 
\centering
\caption{
Parameters of time series $X_L(j)$, $X(j)$, $X_{S,L}(j)$, $X_S(j)$ calculated using various estimators: MSD, AVE(H), Hig(H), D(H). Model series generated using parameters $d=0.3$, $\alpha=1.8$, $\lambda=0.3$, $N=200000$. }

\setlength{\tabcolsep}{3.07mm}
{\begin{tabular}{cccccc}
\toprule
\multicolumn{1}{c}{\textbf{Series}} & \multicolumn{1}{c}{\boldmath{$d$}} & \multicolumn{1}{c}{\boldmath{\textbf{MSD}}} & \multicolumn{1}{c}{\boldmath{\textbf{AVE(H)}}} & \multicolumn{1}{c}{\boldmath{\textbf{Hig(H)}}} & \multicolumn{1}{c}{\boldmath{$D(H)$}} \\ \midrule

$X_L(j)$ & $0.3$ & $1.61$ & $0.84$ & $0.84$ & $0.82\div0.84$  \\ \hline
$X(j)$ & $0.3$ & $1.03$ & $0.25$ & $0.27$ & $0.6\div0.25$  \\ \hline
$X_{S,L}(j)$   & $0.3$ & $1.53$ & $0.81$ & $0.82$ & $0.75\div0.81$  \\ \hline
$X_S(j)$   & $0.3$ & $1.02$ & $0.22$ & $0.24$ & $0.34\div0.16$  \\ 
\bottomrule
\end{tabular}}
\label{table1}
\end{table}
In Table \ref{table1}, we list Hurst parameter evaluation results using different methods for the model series: $X_L(j)$, $X(j)$, $X_{S,L}(j)$, $X_S(j)$. See more detailed information in \cite{Gontis2022CNSNS,Gontis2023FractFrac} about the estimation of mean square displacement (MSD) and $H$ using the Absolute value estimator (AVE) or Higuchi's method. It shows us that formally evaluated Hurst parameters can give us misleading results defining persistence and long-range dependence. Though all series are generated with the same memory parameter $d$, the correct interpretation of self-similarity is compulsory for understanding the memory effects in these time series.

In conclusion, the artificial order disbalance and/or opinion dynamics time series model provides valuable insights into the persistence and memory properties of the limit order flow in financial markets. The comparison with empirical data demonstrates the usefulness of the model and supports the conclusion that the $q$-exponential nature of limit order cancellation times contributes to the observed persistence in order disbalance time series.

\section{Discussion and~Conclusions  \label{sec:Conclusions}}

In our previous work \cite{Gontis2023FractFrac}, we have introduced the concept of a discrete $q$-exponential distribution, see Equation~\eqref{Eq:PMFfromSF}, as a $q$-extension of the geometric distribution, based on the theoretical foundations of generalized Tsallis statistics~\cite{Tsallis1988-ku}. This distribution was acceptable for the fit of empirical limit order cancellation times, revealing their weak sensitivity to order sizes and price levels. The fitted discrete $q$-exponential PMF with parameter $q=1.5$ has proven consistent across ten stocks and trading days analyzed. From the equivalence of $q$-exponential distribution to the second class Pareto distribution \cite{delaBarra2021EPJB}, we know that these distributions have a power-law tail with the exponent $\kappa=\frac{1}{q-1}=2$. We propose the heterogeneous agent model to derive this unique power-law with exponent $\kappa=2$ in this contribution.

We base our idea on ranking trading agents according to their activity during selected time intervals, such as one day. Thus, one can assume having $n$ categories of agents with $i=\{1,2,...,i,..,n\}$ limit orders submitted. Since every order is canceled or executed, it is natural to assume that Geometric PMF $P_i(k) = (1 - \eta i/n)^(k-1) \eta i/n$ describes the lifetime $k$ of agent's group $i$ limit orders. If a number of agents in every group $i$ is inversely proportional to the group's index, Zipf's law, probabilities $P_i(k)$ have the same weight averaging waiting times through the whole set of limit orders. Finally, we get the explicit form PMF of the cancellation (waiting) time \eqref{Eq:PMF-ensemble}, which explains the empirically defined power-law property of limit order cancellation times \cite{Gontis2023FractFrac}.

We generalize the combination of two independent random sequences, ARFIMA\{0,d,0\}\{a, N\} and $P_{\lambda,q}(k)$ Eq. \eqref{Eq:PMFfromSF}, into disbalance series $X(j)$ as the model of opinion dynamics and investigate its long-range dependence. Indeed, the model, first of all, helps understand the properties of limit order disbalance in the financial markets as originates from the analysis of empirical time series \cite{Gontis2020JStat,Kazakevicius2021Entropy,Gontis2022CNSNS}. A more general interpretation of the proposed double stochastic time series might help understand the complexity of long-range dependence in other social systems and empirical time series.

The proposed model serves as an example of a time series with hidden long-range dependence. Thus, we propose the method of self-similarity tests and demonstrate that series $X(j)$ and $X_S(j)$ are not self-similar. Though the result is predictable, the proposed method might be useful in analyzing other empirical time series before using widely accepted methods of self-similar series analysis.

Our study contributes to a better understanding of order disbalance time series and their memory effects in financial markets. Furthermore, the combination of FLSM and the $q$-exponential distribution proves to be a promising approach for modeling social systems, which can be explored further in future research. In conclusion, by bridging the gap between theory and empirical observations, we contribute to developing more accurate models and deeper insights into the behavior of financial markets and social systems.

\section{Abbreviations}
The following abbreviations are used in this manuscript:\\
 
\noindent 
\begin{tabular}{@{}ll}
ARFIMA & Auto--regressive fractionally integrated moving average\\
AVE & Absolute Value estimator\\
FBM & Fractional Brownian motion\\
FGN & Fractional Gaussian noise\\
FLSM & fractional L\`{e}vy stable motion\\
MSD & Mean squared displacement\\
PDF & Probability density function\\
PMF & Probability mass function
\end{tabular}


\end{document}